\documentclass[useAMS,usenatbib]{mn2e}
\usepackage{natbib}
\usepackage{graphicx}
\usepackage{times}
\usepackage{epstopdf}
\usepackage{color}
\usepackage[section]{placeins}
\usepackage[english]{babel}
\usepackage{lineno}
\usepackage{url}
\usepackage{ulem}
\usepackage[colorlinks,linkcolor={blue},citecolor={blue},urlcolor={blue}]{hyperref}
\def\apj{ApJ}

\def\apjl{ApJL}

\def\mnras{MNRAS}
\def\aj{AJ}

\def\xioneh{\xi_{\rm gg}^{\rm 1h}}
\def\xitwoh{\xi_{\rm gg}^{\rm 2h}}
\def\fcs{f_{\rm cs}}
\def\fss{f_{\rm ss}}
\def\xihhcc{\xi_{\rm hh,cc}}
\def\xihhcs{\xi_{\rm hh,cs}}
\def\xihhss{\xi_{\rm hh,ss}}
\def\ni{\bar{n}_i}
\def\nj{\bar{n}_j}
\def\ng{\bar{n}_g}
\def\meanNsati{\langle N_{\rm sat} (M_i)\rangle}
\def\meanNsatj{\langle N_{\rm sat} (M_j)\rangle}
\def\meanNceni{\langle N_{\rm cen} (M_i)\rangle}
\def\meanNcenj{\langle N_{\rm cen} (M_j)\rangle}
\def\meanNsatNsatmone{\langle N_{\rm sat}(M_i) [N_{\rm sat}(M_i)-1]\rangle}
\def\meanNcenNsat{\langle N_{\rm cen}(M_i) N_{\rm sat}(M_i)\rangle}
\def\vecr{{\mathbf r}}
\def\Mi{M_i}
\def\Mj{M_j}

\def\nijg{\frac{\ni\nj}{\ng^2}}

\def\meanNcen{\langle N_{\rm cen} (M)\rangle}
\def\meanNsat{\langle N_{\rm sat} (M)\rangle}
\def\Mmin{M_{\rm min}}
\def\siglgM{\sigma_{\log M}}
\def\hinvMsun{h^{-1}M_\odot}
\def\hinvMpc{h^{-1}{\rm Mpc}}
\def\nc{\bar{n}_c}
\def\ns{\bar{n}_s}
\def\fsat{f_{\rm sat}}

\def\xiss{\xi_{\rm ss}}

\def\xigg{\xi_{\rm gg}}
\def\ximm{\xi_{\rm hh}}
\def\xims{\xi_{\rm hs}}
\def\xiss{\xi_{\rm ss}}
\def\nmaini{\bar{n}_{{\rm h},i}}
\def\nmainj{\bar{n}_{{\rm h},j}}
\def\nsubi{\bar{n}_{{\rm s},i}}
\def\nsubj{\bar{n}_{{\rm s},j}}
\def\psati{p_{\rm sat} (M_i)}
\def\psatj{p_{\rm sat} (M_j)}
\def\pceni{p_{\rm cen} (M_i)}
\def\pcenj{p_{\rm cen} (M_j)}
\def\nmmijg{\frac{\nmaini\nmainj}{\ng^2}}
\def\nmsijg{\frac{\nmaini\nsubj}{\ng^2}}
\def\nssijg{\frac{\nsubi\nsubj}{\ng^2}}

\begin{document}
\title[Galaxy Clustering Modelling with Simulations]{Accurate and Efficient Halo-based Galaxy Clustering Modelling with
Simulations}
\author[Zheng Zheng and Hong Guo]{
Zheng Zheng$^{1}$\thanks{E-mail: zhengzheng@astro.utah.edu} and
Hong Guo$^{2,1}$\\
$^{1}$ Department of Physics and Astronomy, University of Utah, 115 South
1400 East,
Salt Lake City, UT 84112, USA\\
$^{2}$ Shanghai Astronomical Observatory, Chinese Academy of Sciences, Shanghai 200030, China\\
}

\maketitle

\begin{abstract}
Small- and intermediate-scale galaxy clustering can be used to establish the
galaxy-halo connection to study galaxy formation and evolution and to
tighten constraints
on cosmological parameters. With the increasing precision of galaxy
clustering measurements from ongoing and forthcoming large galaxy
surveys, accurate models are required to interpret the data and extract
relevant information. We introduce a method based on high-resolution $N$-body
simulations to accurately and efficiently model the galaxy two-point
correlation functions (2PCFs) in projected and redshift spaces. The basic
idea is to tabulate all information of haloes in the simulations necessary
for computing the galaxy 2PCFs within the framework of halo occupation
distribution or conditional luminosity function. It is equivalent to
populating galaxies to dark matter haloes and using the mock 2PCF
measurements as the model predictions. Besides the accurate 2PCF
calculations, the method is also fast and therefore enables an efficient
exploration of the parameter space. As an example of the method, we decompose
the redshift-space galaxy 2PCF into different components based on the type of
galaxy pairs and show the redshift-space distortion effect in each component.
The generalizations and limitations of the method are discussed.
\end{abstract}


\section{Introduction}

Over the past two decades, large galaxy redshift surveys, such as the
Sloan Digital Sky Survey (SDSS; \citealt{York00}), the Two-Degree Field Galaxy
Redshift Survey (2dFGRS; \citealt{Colless99}), the SDSS-III
\citep{Eisenstein11}, and the WiggleZ Dark Energy Survey \citep{Blake11},
have enabled us to study in detail the large-scale structure of the universe
probed by galaxies.
Galaxy clustering has become a powerful tool to study galaxy formation and
evolution and to learn about cosmology. An informative way to interpret
galaxy clustering is to link galaxies to the underlying dark matter halo
population, whose formation and evolution are dominated by gravitational
interaction and whose properties are well understood with analytic models and
$N$--body simulations.

The commonly adopted descriptions of the connection between galaxies and dark
matter haloes include the halo occupation distribution (HOD; e.g.
\citealt{Jing98,Peacock00,Seljak00,Scoccimarro01,Berlind02,Berlind03,Zheng05})
and the conditional luminosity function (CLF; e.g. \citealt{Yang03}). The
former specifies the probability distribution of the number of galaxies in a
given sample as a function of halo mass, together with the spatial and
velocity distribution of galaxies inside haloes. The latter specifies the
luminosity distribution of galaxies as a function of halo mass. Given a set
of HOD or CLF parameters, with the halo population for a given cosmological
model, galaxy clustering statistics can be predicted. Such frameworks have
been successfully applied to galaxy clustering data to infer the connection
between galaxy properties and halo mass
\citep[see e.g.][]{Bosch03a,Zehavi05,Zehavi11,Zheng07,Zheng09,Guo14,
Skibba15} and to constrain cosmology \citep[e.g.][]{Bosch03b,Tinker05,
Cacciato13,Reid14}.
In particular, the main
clustering statistic used is the two-point correlation function (2PCF) of
galaxies, which is the focus of this paper as well.

Halo properties, like their mass function and spatial clustering (bias), can
be understood analytically \citep[e.g.][]{Press74,Mo96,Sheth99}, and $N$-body
simulations also enable accurate fitting formulae to be obtained
\citep[e.g.][]{Jenkins01,Tinker08,Tinker10}. Based on
these, analytic models of galaxy 2PCF can be developed. The basic idea is
to decompose the 2PCF into contributions from intra-halo and inter-halo
galaxy pairs. The intra-halo component, or the one-halo term, represents the
highly nonlinear part of the 2PCF. The inter-halo component, or the two-halo
term, can be largely modelled by linear theory. Such analytic models have
the advantage of being computationally inexpensive, and they can be used to
efficiently probe the HOD/CLF and cosmology parameter space. However, as
the precision of the 2PCF measurements in large galaxy surveys continues to
improve, the requirement on the accuracy of the analytic models becomes
more and more demanding. As pointed out in \citet{Zheng04a}, an accurate model
of the galaxy 2PCF needs to incorporate the nonlinear growth of the matter
power spectrum \citep[e.g.][]{Smith03}, the halo exclusion effect, and the
scale-dependent halo bias. In addition, the non-spherical shape of haloes
should also be accounted for \citep[e.g.][]{Tinker05,Bosch13}. These are just
factors to be taken into account in computing the real-space or projected
2PCFs. For redshift-space 2PCFs, more factors come into play. An accurate
analytical description of the velocity field of dark matter haloes in the
nonlinear or weakly nonlinear regime proves to be difficult and complex
\citep[e.g.][]{Tinker07,Reid11,Zu13}. Therefore, an accurate analytic model
of redshift-space 2PCFs on small and intermediate scales is still not within
reach.

The above complications faced by analytic models can all be avoided or
greatly reduced if the 2PCF calculation is directly done with the outputs of
$N$-body simulations. With the simulation, dark matter haloes can be
identified, and their properties (mass, velocity, etc) can be obtained. For a
given set of HOD/CLF parameters, one can populate haloes with galaxies
accordingly (e.g. using dark matter particles as tracers) and form a mock
galaxy catalog. The 2PCFs measured from the mock catalog are then the model
predictions used to model the measurements from observations. Such a method of
directly populating simulations have been developed and applied to model galaxy
clustering data \citep[e.g.][]{White11,Parejko13}. This simulation-based model
is attractive, as more and more large high-resolution $N$-body simulations
emerge. It is also straightforward to implement. Once the mock catalog is
produced, measuring the 2PCFs can be made fast (e.g. with tree code). However,
populating haloes with a given set of HOD/CLF parameters is probably the most
time-consuming step, as one needs to loop over all haloes of interest. In
addition, information of individual haloes and tracer particles is needed,
like their positions and velocities. Even with only a subset of all the
particles in a high-resolution simulation, the amount of data can still be
substantial.

The purpose of this paper is to introduce a method that takes the advantage
of the simulation-based model, but being much more efficient in modelling
galaxy clustering. The main idea is to decompose the galaxy 2PCFs and
compress the information in the simulation by tabulating relevant
clustering-related quantities of dark matter haloes. We also apply a similar
idea to extend the commonly used sub-halo abundance matching method (SHAM;
e.g. \citealt{Conroy06}).

The paper is structured as follows. In Section~\ref{sec:method}, we
formulate the method, within the HOD/CLF-like framework and within the
halo/sub-halo framework. In Section~\ref{sec:application}, we show an example
of modelling redshift-space 2PCFs, which also provides an understanding of
the three-dimensional (3D) small- and intermediate-scale galaxy redshift-space
2PCF and its multipoles by decomposing them into the various components.
In Section~\ref{sec:discussion}, we summarize the method and discuss possible
generalizations and limitations.

\section{Simulation-based Method of Calculating Galaxy 2PCFs}
\label{sec:method}

In our simulation-based method, we divide haloes identified in $N$-body
simulations into narrow bins of a given property, which determines galaxy
occupancy. In the commonly used HOD/CLF, the property is the halo mass. In
our presentation, we use halo mass as the halo variable, but the method can
be generalized to any set of halo properties.

The basic idea of the method is to decompose the galaxy 2PCF into
contributions from haloes
of different masses, from one-halo and two-halo terms, and from different
types of galaxy pairs (e.g. central-central, central-satellite, and
satellite-satellite pairs). The decomposition also allows the separation
between the halo occupation and halo clustering. The former relies on the
specific HOD/CLF parameterization, while the latter can be calculated from
the simulation. The method is to tabulate all relevant information about the
latter for efficient calculation of galaxy 2PCFs and exploration of the HOD/CLF
parameter space.

We first formulate the method in the HOD/CLF framework. We then apply the
similar idea to the SHAM case, which provides a more general SHAM method.

\subsection{Case with Simulation Particles}
\label{sec:particle}

Let us start with a given $N$-body simulation and a given set of HOD/CLF
parameters. To populate galaxies into a halo identified in the simulation, we
can put one galaxy at the halo `centre' as a central galaxy, according to
the probability specified by the HOD/CLF parameters. Halo `centre' should be
defined to reflect galaxy formation physics. For example, a sensible choice
is the position of potential minimum rather than centre of mass. For
satellites, we  can choose particles as tracers. In the usually adopted
models, it is assumed that satellite galaxies follow dark matter particles
inside haloes \citep[e.g.][]{Zheng04a,Tinker05,Bosch13}, rooted in theoretical
basis \citep[e.g.][]{Nagai05}. One can certainly modify the distribution
profile as needed, and below we assume that the distribution of galaxies
inside haloes has been specified and that the corresponding tracer particles
have been selected for each halo.

We divide haloes in the simulation into $N$ narrow mass bins and denote the mean
number density of haloes in the mass bin $\log M_i\pm d\log M_i/2 $ as $\ni$.
The mean number density of galaxies is computed as
\begin{equation}
\label{eqn:ng}
\ng=\sum_i\ni[\meanNceni+\meanNsati],
\end{equation}
where $N_{\rm cen}(M)$ and $N_{\rm sat}(M)$ are the occupation numbers of
central and satellite galaxies in a halo of mass $M$, $\langle\rangle$
denotes the average over all haloes of this mass, and $i=1$, ..., $N$.

In the halo-based model, galaxy 2PCF $\xigg$ is computed as the combination
of two terms, $\xigg=1+\xioneh+\xitwoh$ \citep{Zheng04a}, where the one-halo
term $\xioneh$ (two-halo term $\xitwoh$) are from contributions of intra-halo
(inter-halo) galaxy pairs.
Following \citet{Berlind02}, the one-halo term can be computed based on
\begin{equation}
\frac{1}{2} \ng (\ng  d^3\vecr) \left[1+\xioneh(\vecr)\right]
= \sum_i \ni \langle N_{\rm pair}(M_i)\rangle f(\vecr; M_i) d^3\vecr.
\label{eqn:npair1h}
\end{equation}
The left-hand side (LHS) is the number density of one-halo pairs with
separation in the range $\vecr \pm d\vecr/2$ from the definition of 2PCF. The
right-hand side (RHS) is the same quantity from counting one-halo pairs in
each halo and the summation is over all the halo mass bins. Here $\langle
N_{\rm pair}(M)\rangle$ is the total mean number of galaxy pairs in haloes of
mass $M$, and $f(\vecr; M)$ is the probability distribution of pair separation
in haloes of mass $M$, i.e. $f(\vecr; M) d^3\vecr$ is the probability of
finding pairs with separation in the range $\vecr \pm d\vecr/2$ in haloes of
$M$. By further decomposing pairs into central-satellite (cen-sat) and
satellite-satellite (sat-sat) pairs, we reach the following expression,
\begin{eqnarray}
\label{eqn:1h}
1+\xioneh(\vecr) =
\sum_i 2\frac{\ni}{\ng^2} \meanNcenNsat \fcs(\vecr;\Mi) \nonumber \\
+
\sum_i\frac{\ni}{\ng^2} \meanNsatNsatmone \fss(\vecr;\Mi).
\end{eqnarray}
The functions $\fcs(\vecr;M)$ and $\fss(\vecr;M)$ are the probability
distributions of one-halo cen-sat and sat-sat galaxy pair separation in
haloes of mass $M$. They are normalized such that
\begin{equation}
\int \fcs(\vecr;M) d^3\vecr=1 \,\,\,\,{\rm and}\,
\int \fss(\vecr;M) d^3\vecr=1.
\end{equation}
Note that here and in what follows, the 2PCF can be either real-space,
projected-space, redshift-space, or it can be the multipoles of the
redshift-space 2PCF. The variable $\vecr$ should be understood as pair
separation in the corresponding space. For redshift-space clustering,
we discuss how to
specify velocity distribution of galaxies later.

To compute the two-halo term, we add up all possible two-halo galaxy pairs,
following the 2PCF decomposition from different pair counts in \citet{Zu08}.
Similar to equation~(\ref{eqn:npair1h}), the total number density of
two-halo pairs with separation in the range $\vecr \pm d\vecr/2$ is
\begin{equation}
n_{\rm pair,2h} = \frac{1}{2} \ng (\ng  d^3\vecr) \left[1+\xitwoh(\vecr)\right],
\end{equation}
which is composed of two-halo central-central (cen-cen) pairs
\begin{eqnarray}
n_{\rm cc-pair,2h} = &&
\frac{1}{2} \sum_{i\neq j} [\ni\meanNceni][\nj\meanNcenj d^3\vecr]
\nonumber\\
&& \times [1+\xihhcc(\vecr ; \Mi,\Mj)],
\label{eqn:ccpair2h}
\end{eqnarray}
two-halo cen-sat pairs
\begin{eqnarray}
n_{\rm cs-pair,2h} = &&
\sum_{i\neq j}  [\ni\meanNceni][\nj\meanNsatj d^3\vecr]
\nonumber\\
&& \times [1+\xihhcs(\vecr ; \Mi,\Mj)],
\label{eqn:cspair2h}
\end{eqnarray}
and two-halo sat-sat pairs
\begin{eqnarray}
n_{\rm ss-pair,2h} = &&
\frac{1}{2}\sum_{i\neq j}  [\ni\meanNsati][\nj\meanNsatj d^3\vecr]
\nonumber\\
&& \times [1+\xihhss(\vecr ; \Mi,\Mj)].
\label{eqn:sspair2h}
\end{eqnarray}
In each of equations~(\ref{eqn:ccpair2h})--(\ref{eqn:sspair2h}), the summation
is over all halo mass bins (i.e. $i=1$, ..., $N$ and $j=1$, ..., $N$).
The three correlation functions on the RHS have the following meanings --
$\xihhcc(\vecr ; \Mi,\Mj)$ is just the two-point cross-correlation function
between `centres' (positions to put central galaxies) of haloes of masses
$\Mi$ and $\Mj$ (cen-cen);
$\xihhcs(\vecr ; \Mi,\Mj)$ is the two-point
cross-correlation function between the `centres' of $\Mi$ haloes and the
satellite tracer particles in the (extended) $\Mj$ haloes (cen-sat);
$\xihhss(\vecr ; \Mi,\Mj)$ is the two-point cross-correlation function between
satellite tracer particles in the (extended) $\Mi$ haloes and those in the
(extended) $\Mj$ haloes (sat-sat). With $n_{\rm pair,2h}=n_{\rm cc-pair,2h}
+n_{\rm cs-pair,2h} +n_{\rm ss-pair,2h}$, we reach the final expression
for the two-halo term,
\begin{eqnarray}
\label{eqn:2h}
\xitwoh(\vecr)  =
      \sum_{i\neq j}  \nijg \meanNceni\meanNcenj \xihhcc(\vecr ; \Mi,\Mj) \nonumber \\
 +  \sum_{i\neq j} 2\nijg \meanNceni\meanNsatj \xihhcs(\vecr ; \Mi,\Mj) \nonumber \\
 +  \sum_{i\neq j}  \nijg \meanNsati\meanNsatj \xihhss(\vecr ; \Mi,\Mj).
\end{eqnarray}

Equations~(\ref{eqn:ng}), (\ref{eqn:1h}), and (\ref{eqn:2h}) lead to the
method we propose. The quantities related to galaxy occupancy are specified
by the HOD/CLF parameterization one chooses, while those related to haloes
are from the simulation, independent of the HOD/CLF parameterization. We
therefore can prepare tables for $\ni$, $\fcs(\vecr;\Mi)$, $\fss(\vecr;\Mi)$,
$\xihhcc(\vecr ; \Mi,\Mj)$, $\xihhcs(\vecr ; \Mi,\Mj)$, and $\xihhss(\vecr ;
\Mi,\Mj)$. For a given set of HOD/CLF parameters, the predictions of galaxy
2PCFs can be obtained from performing the weighted summation over the tables.
The tables are only prepared once, and we can then change the galaxy
occupation as needed to compute galaxy 2PCFs for different galaxy samples and
different sets of HOD/CLF parameters, which is much more efficient than
populating galaxies into haloes by selecting particles.

Since summation is used to replace integration in the method, we need to
choose narrow halo mass bins ($d\log M=0.01$ is usually sufficient, as shown
in Section~\ref{sec:application}).
The $\ni$ table represents the halo mass function. To prepare
the other tables that depend on pair separation, the bins of pair separation
$\vecr$ are best chosen to match the ones used in the measurements from
observational data, which would naturally avoid any discrepancy related to
the finite bin sizes. For haloes in each mass bin, the $\fcs$ and $\fss$
tables can be computed by using either all the particles in the haloes with
the specified distribution or a random subset. For $\xihhcc$, $\xihhcs$, and
$\xihhss$, we effectively compute the halo-halo two-point cross-correlation
function with different definitions of halo positions.  For $\xihhcc$, halo
positions are defined by our choice of `centres'. For
$\xihhcs(\vecr ; \Mi,\Mj)$, we choose `centres' for $\Mi$ haloes and positions
of arbitrary tracer particles in $\Mj$ haloes. For $\xihhss(\vecr ; \Mi,\Mj)$,
positions of arbitrary tracer particles in both $\Mi$ and
$\Mj$ haloes are chosen. We can use any number of tracer particles in each
halo to do the calculation. For haloes with positions defined by the tracer
particles, they can be thought as extended (with positions having a
probability distribution). On large scales, $\xihhcc$, $\xihhcs$, and
$\xihhss$ are the same, while on small scales, $\xihhcs$ and $\xihhss$ are
smoothed version of $\xihhcc$. Note that in analytic models such
differences are usually neglected. In computing the three halo-halo
correlation functions, we do not need to construct random catalogs to find
out the pair counts from a uniform distribution -- in the volume
$V_{\rm sim}$ of the simulation with periodic boundary conditions, the counts of
cross-pairs at separation in the range $\vecr \pm d\vecr/2$ between two
randomly distributed populations with number densities $\ni$ and $\nj$ are
simply $(\ni V_{\rm sim}) (\nj d^3\vecr)$. Making use of this fact can
greatly reduce the computational expense in preparing the tables.

For the redshift-space tables, in addition to the halo velocities, one needs
to specify the velocity distribution of galaxies inside haloes, which can be
different from that of dark matter particles (a.k.a. velocity bias; e.g.
\citealt{Berlind02}). The difference can be parameterized by central and
satellite velocity bias parameters \citep[e.g.][]{Guo15a}. For a set of
central and satellite velocity bias parameters and with a choice of the
line-of-sight direction, we can obtain the redshift-space positions of the
central galaxy and satellite tracer particles according to halo velocities
and central and satellite galaxy velocity distributions inside haloes, and
the redshift-space tables can be computed. We suggest to prepare tables for
different sets of central and satellite velocity bias parameters and
interpolate among tables to probe the velocity bias
parameter space, as is done in \citet{Guo15a}.

Multipole moments of redshift-space galaxy 2PCFs are usually modelled. We can
derive the corresponding tables by computing the corresponding multipole
moments of $\fcs$, $\fss$, $\xihhcc$, $\xihhcs$, and $\xihhss$. In such a
case, $r$ is expressed by $s=|\vecr|$ and $\mu$, the cosine of the angle
between $\vecr$ and the line-of-sight direction. In the integration
(summation) for obtaining the multipoles, the bins of $\mu$ match those used
in observational measurements to remove any finite-bin-size effect.

For modelling the projected 2PCF $w_p$, a corresponding set of tables can be
obtained by integrating the redshift-space tables over the line-of-sight
separation.
The integration is done in the same way as in the measurements with data to
avoid any finite-bin-size effect, summing over the same line-of-sight bins
(with the same bin size) up to the same maximum line-of-sight separation.

\subsection{Case with Subhaloes}
\label{sec:subhalo}

The SHAM method uses more information from (high-resolution) simulations,
including both distinct haloes and subhaloes identified inside distinct 
haloes, where the distinct haloes refer to haloes that are not subhaloes of 
another halo. Distinct haloes are also referred to as haloes, main haloes, 
or host haloes.  Central galaxies are hosted by distinct haloes at the centres,
while satellite galaxies are in subhaloes. Before merging into distinct
haloes, subhaloes are distinct haloes themselves. The SHAM method generally
works in the following way. By adopting one property, subhaloes and distinct
haloes can be treated as a unified entity. For distinct haloes, the property is
evaluated at the time of interest. For subhaloes, it becomes common practice
to evaluate the property at the time when subhaloes were still distinct haloes.
The properties commonly used include mass ($M_{\rm acc}$) at the time a 
subhalo was accreted into a host halo, maximum circular velocity
$V_{\rm acc}$ at the time of accretion, and peak maximum circular velocity
$V_{\rm peak}$ over the history of the subhalo as a distinct halo.
The connection between haloes/subhaloes and galaxies is established by rank
ordering haloes/subhaloes according to the given property and galaxies
according to one certain property (e.g. luminosity or stellar mass). When
normalized to the same survey/simulation volume, halo/subhalo and galaxy of
the same rank are linked. A more general treatment also accounts for the 
scatter between the halo/subhalo property and the galaxy property. The 
simple procedure of linking light (galaxies) to matter (haloes/subhaloes) 
can provide a reasonable interpretation of galaxy clustering trend and enable 
a study of galaxy evolution \citep[e.g.][]{Conroy06,Conroy09,Behroozi13,Reddick13}.

We generalize the idea in Section~\ref{sec:particle} to the subhalo case,
extending the SHAM model and making it efficient to model galaxy clustering.
The model allows the scatter between the halo/subhalo property and the galaxy
property to be different for distinct haloes (central galaxies) and subhaloes
(satellite galaxies). We use mass as the halo/subhalo property variable here,
which can be understood as the mass at accretion ($M_{\rm acc}$). However, it
can be replaced by any property one chooses to adopt, e.g. $V_{\rm acc}$ and
$V_{\rm peak}$. A halo/subhalo method following a similar spirit of pair
decomposition to model the projected galaxy 2PCF and weak lensing signal is
presented in \citet{Neistein11} and \citet{Neistein12}.

Compared to the commonly used SHAM method that connects the whole range of
galaxy property and halo/subhalo property, the method presented here works for
each individual galaxy sample. To some degree, it is formulated in an
HOD/CLF-like form, with distinct haloes and subhaloes as tracers of central
and satellite galaxies, respectively. It is no longer limited to 
{\it abundance} matching. Instead, the method can be used to fit 
both galaxy abundance and galaxy clustering (2PCFs). 

For a given galaxy sample, the scatter between halo/subhalo property and
galaxy property means that not all haloes/subhaloes are fully occupied by
these galaxies, which can be characterized by the probability of occupancy
(or the smaller-than-unity mean occupation number). Denote the mean
occupation number of central galaxies in distinct haloes of mass $M_h$ as
$p_{\rm cen}(M_h)$ and that of satellite galaxies in subhaloes of mass $M_s$
as $p_{\rm sat}(M_s)$. The same bins of mass are adopted for $M_h$ and $M_s$. In
principle we do not need to differentiate $M_s$ and $M_h$, since the scripts
of `c' (cen) and `s' (sat) below make the situation self-explanatory. Let
the mean number densities of distinct haloes and subhaloes in the mass bin
$\log M_i\pm d\log M_i/2 $ be $\nmaini$ and $\nsubi$, respectively.

For a given sample of galaxies, with a model of $p_{\rm cen}(M)$ and $p_{\rm
sat}(M)$, the mean number density of galaxies $\ng$ is computed as
\begin{equation}
\ng=\sum_i[\nmaini\pceni+\nsubi\psati].
\end{equation}
With a similar decomposition as in equation~(\ref{eqn:2h}), the galaxy 2PCF
can be computed as
\begin{eqnarray}
\xigg(\vecr)  =
      \sum_{i,j}  \nmmijg \pceni\pcenj \ximm(\vecr ; \Mi,\Mj) \nonumber \\
 +  \sum_{i,j} 2\nmsijg \pceni\psatj \xims(\vecr ; \Mi,\Mj) \nonumber \\
 +  \sum_{i,j}  \nssijg \psati\psatj \xiss(\vecr ; \Mi,\Mj),
\end{eqnarray}
which simply states that the total number of galaxy pairs is the sum of
cen-cen, cen-sat, and sat-sat pairs.
The three correlation functions on the RHS have the following meanings --
$\ximm(\vecr ; \Mi,\Mj)$ is just the two-point cross-correlation function
between centres of
distinct haloes of masses $\Mi$ and $\Mj$; $\xims(\vecr ; \Mi,\Mj)$ is the
two-point cross-correlation function between centres of $\Mi$ distinct haloes
and those of $\Mj$ subhaloes; $\xiss(\vecr ; \Mi,\Mj)$ is the two-point
cross-correlation function between centres of subhaloes of masses $\Mi$ and
$\Mj$. Unlike the particle case in Section~\ref{sec:particle}, there are no
explicit one-halo and two-halo terms here (though they can be derived), and
the $i\neq j$ condition is not imposed in the summation.

The quantities $p_{\rm cen}(M)$ and $p_{\rm sat}(M)$ come from the occupation
function model, which is up to our choice of parameterization for the sample
of galaxies. In this halo/subhalo-based method, we only need to prepare tables
for $\nmaini$, $\nsubi$, $\ximm(\vecr ; \Mi,\Mj)$, $\xims(\vecr ; \Mi,\Mj)$,
and $\xiss(\vecr ; \Mi,\Mj)$.

As with the tables using particles (Section~\ref{sec:particle}), for
redshift-space 2PCF or multipole moments, tables for different sets of
central and satellite velocity bias parameters can be prepared. For each set,
haloes and subhaloes are shifted to redshift-space positions for calculation.
Tables can also be generated for modelling the projected 2PCF $w_p$.
The procedures and bins used in the measurements should be followed so that
the model and measurements are made fully consistent.

\section{An Example Application and the Redshift-Space 2PCF Decomposition}
\label{sec:application}

\begin{figure*}
\includegraphics[width=0.9\textwidth]{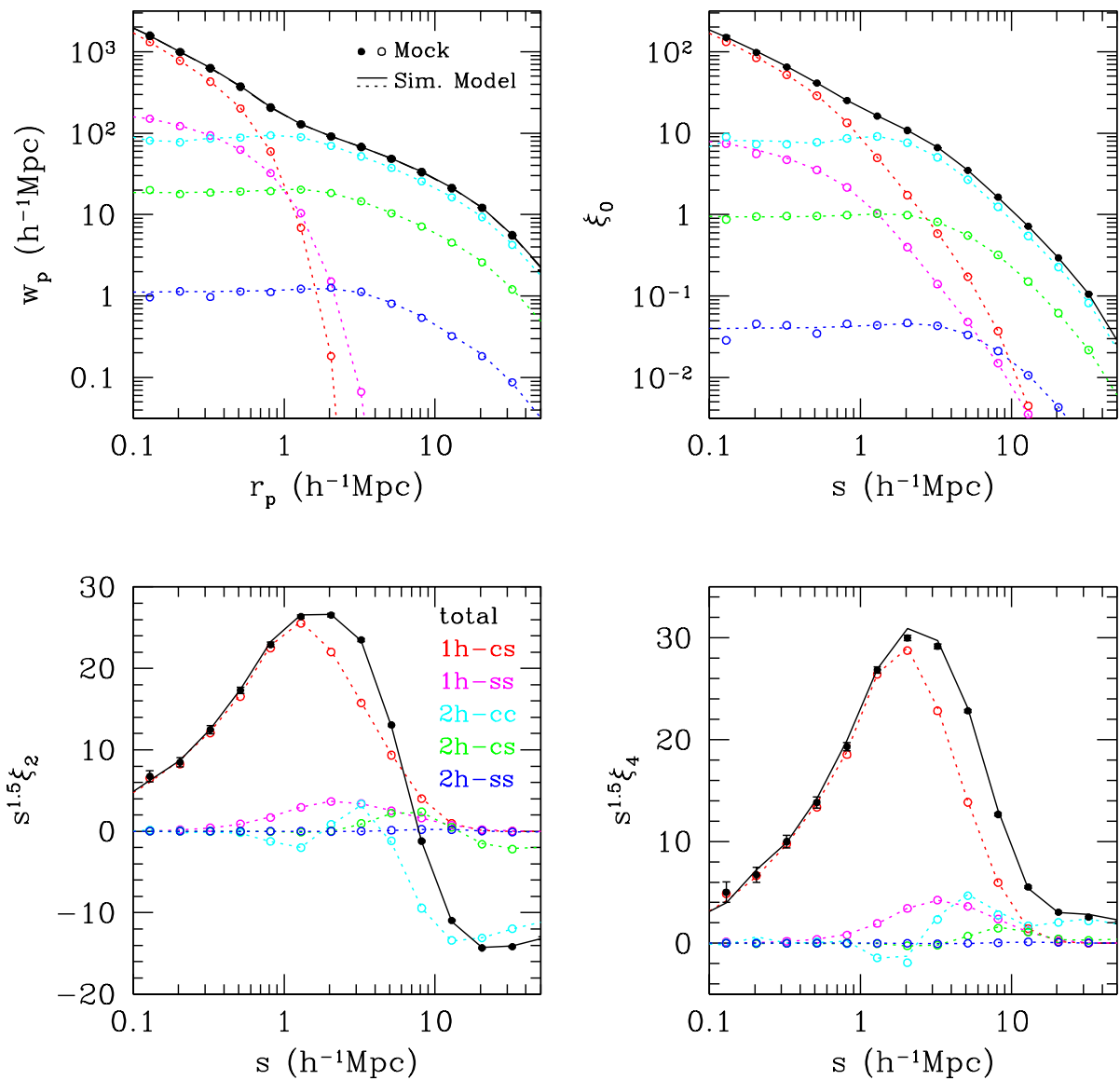}
\caption{ \label{fig:wpxi024} Decomposition of the projected galaxy 2PCF
$w_p$ and redshift-space 2PCF multipoles $\xi_0$, $\xi_2$, and $\xi_4$ into
the various one-halo and two-halo components (one-halo cen-sat, one-halo
sat-sat, two-halo cen-cen, two-halo cen-sat, and two-halo sat-sat). The
circles are measurements from 100 mock galaxy catalogs constructed by
populating galaxies into dark matter halos in the simulation, according to
the set of fiducial HOD parameters. The curves are calculations with the
method introduced in this paper. See text for more details. }
\end{figure*}

\begin{figure*}
\includegraphics[width=1.0\textwidth]{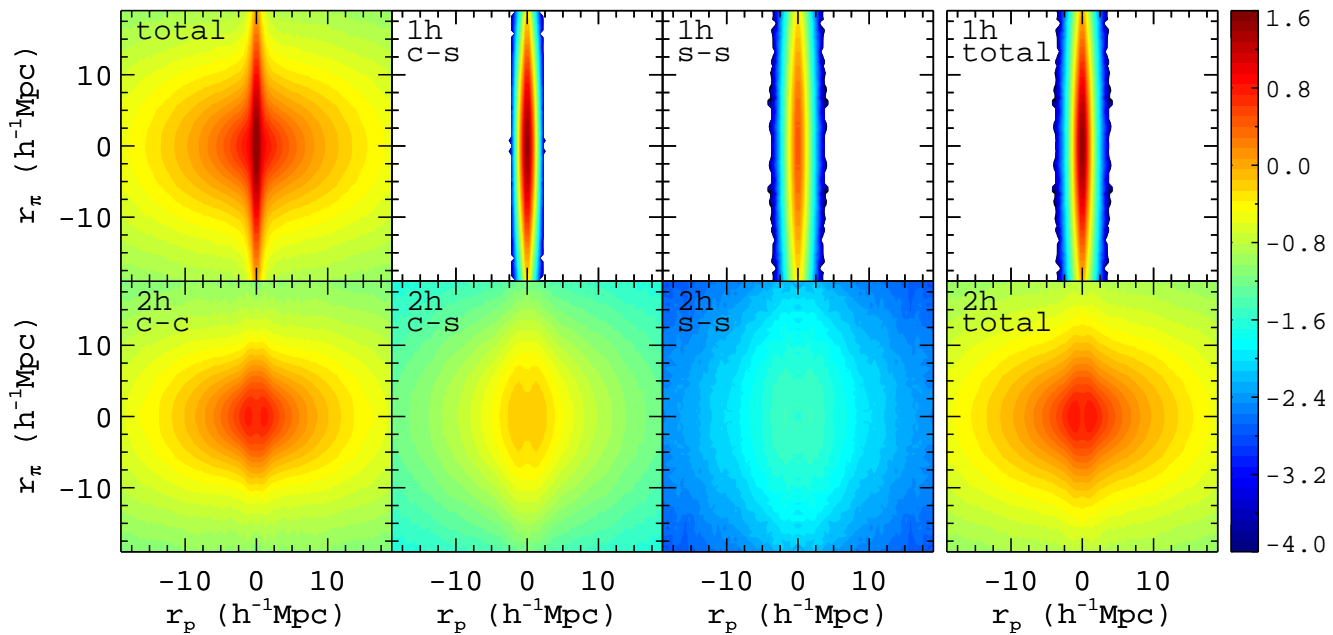}
\caption{ \label{fig:xi3d} Decomposition of the 3D redshift-space 2PCF
$\xi(r_p,r_\pi)$ into the various one-halo and two-halo components (one-halo
cen-sat, one-halo sat-sat, two-halo cen-cen, two-halo cen-sat, and two-halo sat-sat).
The plot is based on the average measurements from 100 mock galaxy catalogs
constructed by populating galaxies into dark matter halos in the simulation,
according to the set of fiducial HOD parameters. The color scale shows
$\xi(r_p,r_\pi)$ in logarithmic scale. See text for more details. }
\end{figure*}

The method developed here has been successfully applied to model projected
and redshift-space 2PCFs of SDSS and SDSS-III galaxies on small to
intermediate scales (e.g. \citealt{Guo15a,Guo15b,Guo15c}) and to compare
HOD and SHAM models (Guo et~al. in prep.).
As the method is built on the basis of decomposition of galaxy 2PCFs,
here we provide an example to illustrate the different 2PCF components. In
particular, we show the components for the redshift-space 3D 2PCF and the
manifestation of redshift-space distortions in each component to have a better
understanding of the redshift-space 2PCFs within the HOD framework. In
addition, we also investigate how redshift-space 2PCFs help with HOD
constraints, including the inference of the galaxy velocity distribution
inside haloes.

The example adopts HOD parameters for the sample of $z\sim 0.5$ CMASS
galaxies in the the SDSS-III Baryon Oscillation Spectroscopic Survey
(BOSS; \citealt{Dawson13}). With spherical overdensity haloes and halo
particles from the $z=0.53$ output of the MultiDark simulation (MDR1;
\citealt{Prada12,Riebe13}), we create tables for halo properties, including
halo number density $\bar{n}$ (i.e. halo mass function), projected 2PCF $w_p$,
redshift-space 2PCF monopole $\xi_0$, quadrupole $\xi_2$, and hexadecapole
$\xi_4$. We choose the position of the potential minimum as the centre
of each halo for putting the central galaxy and halo particles as tracers of
satellites. Each of $w_p$ and $\xi_{0/2/4}$ has five components (one-halo
cen-sat, one-halo sat-sat, two-halo cen-cen, two-halo cen-sat, and two-halo
sat-sat). To generate the $w_p(r_p)$ tables, we measure $\xi(r_p,r_\pi)$ for
each component and for each combination of halo mass bins and sum over the
$r_\pi$ direction, where $r_p$ and $r_\pi$ are the pair separations in the
directions perpendicular and parallel to the line-of-sight direction (chosen
to be one principle direction of the simulation box). To generate the
$\xi_{0/2/4}$ tables, we measure $\xi(s,\mu)$ for each component and for each
combination of halo mass bins and form the multipoles by integrating over
$\mu$, where $s$ is the redshift-space pair separation and $\mu$ the cosine
of the angle between pair displacement and the line-of-sight direction.
Following the setup in the observational measurements \citep{Guo15a}, we have
19 bins for $r_p$ and $s$ uniformly spaced in logarithmic space, 50 linearly
spaced bins in $r_\pi$ and 20 linearly spaced bins in $\mu$. For halo mass
bins, we use $d\log M=0.01$. We construct tables for 5 bins of central
velocity bias parameter $\alpha_c$ and 8 bins of satellite velocity bias
parameter $\alpha_s$, respectively. The total size of the final set of tables
is about 10GB. That is, the information in the high-resolution simulation
output relevant for modelling projected and redshift-space 2PCFs of galaxies
has been tremendously compressed, making the modelling tractable even with
a desktop computer.

For the HOD, we adopt the common parameterization for a sample of galaxies
above a luminosity threshold \citep{Zheng05,Zheng07}. The mean occupation
function of central galaxies in haloes of mass $M$ is
\begin{equation}
\meanNcen = \frac{1}{2}\left[1+{\rm erf}\left(\frac{\log M -\log \Mmin}{\siglgM}\right)\right],
\end{equation}
where ${\rm erf}$ is the error function. For the mean occupation function of
satellite galaxies, we use
\begin{equation}
\meanNsat = \meanNcen \left(\frac{M-M_0}{M_1^\prime}\right)^\alpha.
\end{equation}
The number of satellites in haloes of mass $M$ is assumed to follow the
Poisson distribution with the above mean. In addition, for modelling
redshift-space 2PCFs, we have two additional HOD parameters $\alpha_c$ and
$\alpha_s$ for central and satellite velocity bias. Essentially, $\alpha_c$
($\alpha_s$) is the ratio of the velocity dispersion of central (satellite)
galaxies to that of dark matter particles inside halos (see \citealt{Guo15a}).
For the fiducial model, we
adopt the set of parameters that fit the projected and redshift-space 2PCFs
for the CMASS sample in \citet{Guo15a} -- $\log\Mmin=13.36$, $\siglgM=0.64$,
$\log M_0=13.20$, $\log M_1^\prime=14.23$, $\alpha=1.05$, $\alpha_c=0.30$,
and $\alpha_s=0.91$. Halo masses are in units of $\hinvMsun$.

With the tables and the fiducial HOD parameters, we follow
equations~(\ref{eqn:ng}), (\ref{eqn:1h}), and (\ref{eqn:2h}) to compute all
the components of $w_p$ and $\xi_{0/2/4}$. For the purpose of a sanity check,
we also measure the components from 100 mock galaxy catalogs. The mock
catalogs are generated from populating haloes in the simulation by putting
central galaxies at the potential minimum in haloes and drawing
random dark matter particles as satellite galaxies, in accordance with the
occupation distributions and velocities set by the fiducial HOD parameters.
For the purpose of comparison with the model based on the tables, we
decompose the galaxy 2PCF (either $w_p$ or $\xi_{0/2/4}$) measured in the
mock catalogs into five components,
\begin{eqnarray}
\label{eqn:xigg}
\xigg(\vecr) & = &  2\frac{\bar{n}_{\rm cs-pair}}{\ng^2}f_{\rm cs}(\vecr)
              + 2\frac{\bar{n}_{\rm ss-pair}}{\ng^2}f_{\rm ss}(\vecr)\nonumber\\
             &   &
              + \frac{\nc^2}{\ng^2} \xi_{\rm cc}(\vecr)
              + 2\frac{\nc\ns}{\ng^2} \xi_{\rm cs}(\vecr)
              + \frac{\ns^2}{\ng^2} \xi_{\rm ss}(\vecr).
\end{eqnarray}
The first two terms on the RHS are one-halo terms -- $\bar{n}_{\rm cs-pair}$
and $\bar{n}_{\rm ss-pair}$ are the mean number densities of one-halo
cen-sat pairs and one-halo sat-sat pairs measured in the
mock catalogs, and $f_{\rm cs}$ and $f_{\rm ss}$ are the normalized average
distributions of one-halo cen-sat and sat-sat pairs in
the mock. The last three terms on the RHS are two-halo terms -- $\nc$ and
$\ns$ are the mean number densities of central and satellite galaxies in the
mock, and $\xi_{\rm cc}$, $\xi_{\rm cs}$, $\xi_{\rm ss}$ are the 2PCFs by
counting only two-halo cen-cen, cen-sat, and sat-sat pairs \citep{Zu08}.

Figure~\ref{fig:wpxi024} shows the decomposition of $w_p$ and $\xi_{0/2/4}$
for the fiducial model. As expected, the calculations from the
simulation-based method (curves) agree with the measurements from the mock
catalogs (circles), which is reassuring. For the projected 2PCF $w_p$
(top-left panel), the one-halo cen-sat term (red) dominate the
small-scale signal. The one-halo sat-sat term (magenta) extends to
larger scales, since the maximum sat-sat pair separation in a halo is the
diameter of the halo, twice that of the cen-sat pair separation. Owing to the
low satellite fraction ($\fsat\sim 7\%$) of this sample of galaxies, the
contribution of the one-halo sat-sat pairs to $w_p$ is overall small, but
noticeable around $1\hinvMpc$, the one-halo to two-halo term transition scales.
On large scales, the three two-halo terms have a similar shape, since they
essentially follow the halo-halo correlation. The flattening towards small
scales are caused by the halo exclusion effect. Compared to the two-halo
cen-cen component, the two-halo cen-sat is smoothed on small scales, since each
halo contributing the satellite of the cen-sat pair on average is extended
instead of a point source (the case for the halo contributing the central
galaxy of the pair) as a result of the spatial distribution of satellites
inside haloes. The two-halo sat-sat term is even more smoothed, since every
halo becomes extended. To see the relative contribution of each term to the
large-scale 2PCF, we note that in equation~(\ref{eqn:xigg}), $\xi_{\rm
cc}\propto b_c^2$, $\xi_{\rm cs}\propto b_cb_s$, and $\xi_{\rm ss}\propto
b_s^2$ on large scales, where $b_c$ and $b_s$ are the large-scale bias
factors for central and satellite galaxies, respectively. Since satellites on
average reside in more massive haloes than central galaxies, the value of
$b_s$ is higher than that of $b_c$ (roughly by tens of per cent for
luminosity-threshold samples). From equation~(\ref{eqn:xigg}), we see that
the relative contributions to the large-scale 2PCF from the two-halo cen-cen,
cen-sat, and sat-sat terms are 1 : $2f_nf_b$ : $(f_nf_b)^2$, with
$f_n=\ns/\nc=\fsat/(1-\fsat)$ the satellite to central galaxy number density
ratio and $f_b=b_s/b_c$ the satellite to central galaxy bias ratio. For the
sample we consider, the ratios are 1: 25\% : 1.6\%. For lower luminosity
samples with higher satellite fractions, we expect the contributions from the
two-halo cen-sat and sat-sat to be substantially higher.

The decomposition of the redshift-space 2PCF monopole $\xi_0$ (top-right
panel) and the relative amplitudes of the various terms are similar to the
case of $w_p$. The bottom two panels show the case of quadrupole $\xi_2$ and
hexadecapole $\xi_4$, and a factor $s^{1.5}$ is multiplied for each term so
that both the small-scale and large-scale signals can reasonably show up.
The Fingers-of-God effect \citep{Jackson72,Huchra88} from one-halo terms causes
a positive quadrupole. In the $\xi_2$ panel, we see that the influence of the
one-halo terms can extend to about 10$\hinvMpc$ in the quadrupole. The negative
quadrupole on large scales manifests the Kaiser effect
\citep{Kaiser87,Hamilton92} caused by the coherent motion of haloes, falling
into overdense regions and streaming out of underdense regions. The two-halo
cen-cen term dominates the large-scale quadrupole, but the cen-sat term is
also important. Both terms show low positive quadrupole signals toward small
scales caused by the random motion of haloes (and galaxies). The two-halo
sat-sat term makes an almost negligible contribution to the quadrupole on all
scales. The hexadecapole $\xi_4$ (bottom-right panel) are mostly positive
from all components. The relative contributions from different components are
similar to the quadrupole case.

The projected 2PCF and the redshift-space 2PCF multipoles are usually the
quantities to model. The 3D redshift-space 2PCF measurements are commonly
displayed as contours of $\xi(r_p,r_\pi)$, which make the redshift-space
distortion effects on all scales easily visualized. It would be instructive
to have the corresponding one-halo and two-halo components to gain a better
intuition about the redshift-space distortions. Figure~\ref{fig:xi3d} shows
such a decomposition measured from the mock catalogs, which can also be
calculated using the $\xi(r_p,r_\pi)$ component tables.

\begin{figure*}
\includegraphics[width=0.8\textwidth]{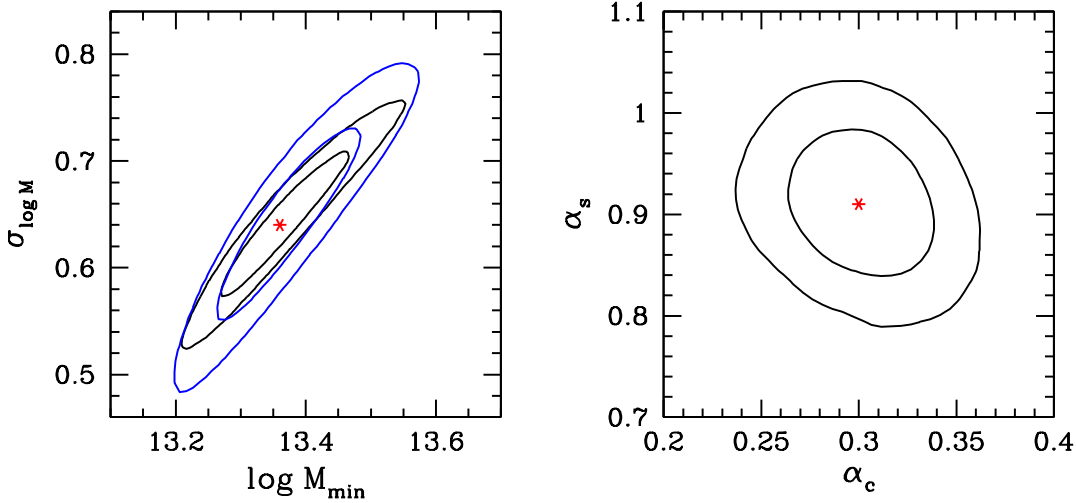}
\caption{ \label{fig:hod} {\it Left:} Constraints on $\log\Mmin$ and
$\siglgM$ from the 2PCFs with the fiducial galaxy sample. The model 2PCFs are
calculated with method introduced in this paper. Blue and black contours are
for the cases of modelling $w_p$ only and jointly modelling
$w_p+\xi_{0/2/4}$, respectively. The 68.3\% and 95.4\% confidence levels are
shown for each case. {\it Right:} Constraints on the central and satellite
velocity bias parameters ($\alpha_c$ and $\alpha_s$) for the fiducial galaxy
sample from jointly modelling $w_p+\xi_{0/2/4}$. The red asterisk in each
panel indicates the value from the fiducial model. }
\end{figure*}
The leftmost panel shows the total redshift-space 2PCF of the sample, with
the Fingers-of-God and Kaiser effects clearly seen. The Fingers-of-God effect,
limited to small transverse separation $r_p$, is mainly contributed by the
one-halo terms (two middle panels on the top). The one-halo sat-sat component
appears to be more extended than the one-halo cen-sat component in both the
transverse and the line-of-sight direction. In the transverse direction, it
can be explained by the fact that the largest one-halo sat-sat (cen-sat) pair
separation is about the diameter (radius) of the largest haloes. In the
line-of-sight direction, the elongation is mainly a result of galaxy motion
inside haloes. The relative line-of-sight velocity of sat-sat pairs are
higher than that of cen-sat pairs, causing the one-halo sat-sat component to be
more extended (shallower profile as a function of $r_\pi$). The total one-halo
term (rightmost panel on the top) is dominated by the cen-sat and sat-sat
component at small $r_p$ and slightly large $r_p$, respectively.

The three two-halo components and the total two-halo term are shown in the
bottom panels of Figure~\ref{fig:xi3d}. In each component, the double-hump
feature at small $r_p$ reflects the halo-exclusion effect. The effect would
lead to a hole at the centre if the real-space 2PCF were plotted here. The
shift in the line-of-sight galaxy positions in redshift space from galaxy
peculiar motion makes the hole partially filled. The two-halo cen-cen
component shows an overall Kaiser squashing effect along the line of sight.
However, the contours at small
$r_p$ are elongated along the line of sight, like the Fingers-of-God effect.
This is caused by the random motion of haloes and that of central galaxies
with respect to haloes (i.e. a non-zero central velocity bias). The two-halo
cen-sat component shows a much stronger line-of-sight elongation up to a few
Mpc in $r_p$. The reason lies in the motion of satellites inside haloes,
which causes the average redshift-space distribution of satellites appears
extended along the line of sight in an average halo hosting the satellites of
the two-halo cen-sat pairs. The line-of-sight elongation pattern is even
stronger in the two-halo sat-sat component -- the correlation of elongated
haloes (as a result of the redshift-space spatial distribution of satellites
inside haloes) completely
suppresses the Kaiser effect even on the largest scales shown here ($\sim
20\hinvMpc$). The total two-halo term is dominated by the cen-cen component
with a substantial contribution from the cen-sat component. The sat-sat
component does not make an important contribution for this sample. As
discussed before, we expect the two-halo cen-sat and sat-sat components to
become more important for galaxy samples with lower luminosity thresholds and
higher satellite fractions.

Overall, for the 3D redshift 2PCF $\xi(r_p,r_\pi)$ different components of
the one-halo and two-halo terms have different transverse range of the
line-of-sight elongation. The profile along the line of sight also depends on
the type of pairs in consideration, becoming increasingly shallower from
cen-cen, cen-sat, to sat-sat components. For each component, the streaming
model \citep[e.g.][]{Peebles80} usually adopted in simple models of
redshift-space distortions should work well, which is kind of a convolution of
the real-space 2PCF with a velocity dispersion kernel. For the total
redshift-space 2PCF, our results indicate that it
is hard to use a single velocity dispersion kernel to accurately model the
redshift-space distortion effect. The different components are needed if one
wishes to develop an accurate analytic model \citep[e.g.][]{Tinker07}.

Finally, we investigate the constraints on the HOD parameters from projected
and redshift-space 2PCFs. The 2PCFs predicted from the fiducial set of HOD
parameters are used as the input measurements, and the full covariance matrix
from \citet{Guo15a} measured from the CMASS data is adopted. The model
uncertainty caused by the finite volume of the simulation is also accounted
for by rescaling the covariance matrix (see Appendix~\ref{sec:appendix}). We
employ a Monte Carlo Markov Chain method to explore the parameter space of
the 7 HOD parameters, $\Mmin$, $\siglgM$, $M_0$, $M_1^\prime$, $\alpha$,
$\alpha_c$, and $\alpha_s$. We first model the projected 2PCF $w_p$ only. The
first five parameters related to the galaxy mean occupation function can be
constrained, while there are virtually no constraints on the velocity bias
parameters ($\alpha_c$ and $\alpha_s$) as the line-of-sight information is
lost. We then jointly model $w_p$ and the redshift-space 2PCF multipoles
$\xi_{0/2/4}$. We find that redshift-space 2PCFs help tighten the constraints
mainly in $\Mmin$ and $\siglgM$, the two parameters for the mean occupation
function of central galaxies. In the left panel of Figure~\ref{fig:hod}, we
compare the constraints (marginalized 1$\sigma$ and 2$\sigma$ contours) from
$w_p$ only (blue) and $w_p$+$\xi_{0/2/4}$ (black). The constraints on the
parameters for the mean occupation function of satellite galaxies are only
slightly improved, mainly in $M_0$. In general, compared to the $w_p$-only
case, redshift-space 2PCFs do not lead to a substantial improvement in the
HOD parameters related to the occupation function. The reason may be related
to the fact that the projected 2PCF $w_p$ is not independent of the
redshift-space 2PCFs, and that the information content in $\xi_{0/2/4}$ to
constrain the occupation-related parameters is largely overlapped with that
in $w_p$. The correlated information in $w_p$ and $\xi_{0/2/4}$ is embedded
in the covariance matrix. Therefore, when jointly modelling $w_p$ and
$\xi_{0/2/4}$, it is important to use the full covariance matrix including
the covariances between $w_p$ and $\xi_{0/2/4}$ to avoid double counting the
information content and artificially tightening the HOD constraints.

The redshift-space distortions are caused by the peculiar motion of galaxies.
The peculiar motion of haloes is in the simulation and built in the tables.
So modelling redshift-space 2PCFs lead to constraints of galaxy motion inside
haloes, i.e. the central and satellite velocity bias parameters. The right
panel of Figure~\ref{fig:hod} shows that velocity bias parameters can be
clearly detected for the fiducial sample. Velocity bias parameters have been
constrained from redshift-space clustering for the $z\sim 0.5$ BOSS CMASS
galaxies \citep{Guo15a,Guo15b,Reid14} and $z\sim 0.1$ SDSS Main galaxies
(see \citealt{Guo15c} and \citealt{Guo15d} for applying the modelling method 
based on simulation particles and subhaloes, respectively). More discussions 
on the velocity bias constraints and the implications can be found in 
\citet{Guo15a}.

\section{Summary and Discussion}
\label{sec:discussion}

In this paper, we introduce a simulation-based method to accurately and
efficiently model galaxy 2PCFs in projected and redshift spaces. The basic
idea is to make use of a high-resolution simulation and tabulate all the halo
information necessary for galaxy clustering calculation. Then on top of the
tables, galaxy 2PCFs can be computed with the galaxy-halo relation specified
by the HOD or CLF model. We also provide a version that applies to and
extends the SHAM method. Based on the method, we also study the decomposition
of the projected and redshift-space galaxy 2PCFs into different components
according to the type of galaxy pairs.

The proposed method is accurate, since it is directly based on
high-resolution simulations. The effects like halo exclusion, nonlinear
evolution, scale-dependent halo bias, and non-sphericity of haloes, which are
difficult to deal with in analytic methods of computing galaxy 2PCFs, are all
automatically accounted for in the simulation-based method. The method also
breaks the 2PCFs into all the one-halo and two-halo components based on the
nature of galaxy pairs and computes each component accurately, which are
usually not the case in analytic methods (especially for the two-halo term).
When building the tables, the same
binning scheme (in pair separation and in angle) and the same integration
procedure as used in the observation measurements are adopted, so there is no
binning-related issue when comparing the model prediction with the
measurements. The method is equivalent to measure the model galaxy 2PCFs from
mock catalogs and is as accurate as what the mean mock catalog can achieve. 
The mock catalogs are constructed by populating galaxies
(using tracer particles) to haloes identified in the simulation, according to
the halo occupation specified by the HOD/CLF model. However, the method is
more efficient, as it avoids the construction of mock catalogs and the
measurement of the 2PCFs from the mocks. Instead, `populating galaxies' and
`measuring the 2PCFs' are performed analytically within the HOD/CLF
framework. This greatly reduces the computational time and make it possible
to efficiently explore the parameter space when modelling the 2PCF data.

A similar method working in Fourier space can be easily developed to model
galaxy redshift-space power spectrum. The method can also be generalized to
other clustering statistics, e.g. angular 2PCF of galaxies, two-point
cross-correlation function of galaxies, and galaxy-galaxy lensing.
Generalizing the method to three-point correlation function (3PCF) of
galaxies is also possible. In principle, there are more components for the
3PCF -- cen-sat-sat and sat-sat-sat triplets for the one-halo term,
cen-(cen-sat), cen-(sat-sat), sat-(cen-sat), and sat-(sat-sat) triplets for
the two-halo term (the pair in the parentheses is in the same halo), and
cen-cen-cen, cen-cen-sat, cen-sat-sat, and sat-sat-sat triplets for the
three-halo term. More importantly, compared to the 2PCF case, the dimension of
each 3PCF component table will increase (e.g. two sides and the angle in
between for a triangle configuration and three halo mass indices). To make
such a method suitable for the 3PCF modelling, further simplification is
necessary, e.g. through multipole or Fourier expansion
\citep[e.g.][]{Szapudi04,Zheng04b,Slepian15}.

To make use of the high precision of small- to intermediate-scale 2PCFs
measurements to help constrain cosmological parameters
\citep[e.g.][]{Zheng07b,Reid14}, a set of tables need to be prepared based on
simulations with different cosmological parameters or by rescaling one 
simulation to different cosmological models \citep[e.g.][]{Zheng02,Tinker06,
Angulo10,Reid14,Guo15c}. Even with one cosmological model, there may be 
situations that need more tables. For example, in the particle-based
model, random particles are selected to trace satellite galaxies by default.
However, the difference between the spatial distributions of satellites and
dark matter can be an additional parameter to be constrained. For such a
purpose, one needs to build different sets of tables using tracer particles
of different distributions. In either of the above cases (or any case that
needs to extend the tables), the total size of the tables would have an
order-of-magnitude increase. Compared with methods of directly populating
simulations, such an increase in table size is still reasonable and manageable.

With one simulation, we do not have the global or ensemble average properties
of haloes. That is, the model with one simulation has uncertainty caused by
the finite volume effect. One can use multiple simulations with different
realizations of the initial conditions to build the average tables, which
reduces the model uncertainty. The model uncertainty should be included in
modelling data. In Appendix~\ref{sec:appendix}, we show that this can be done
by rescaling the covariance matrix of the measurements based on the ratio of
simulation and survey volume. For any simulation, the fluctuation modes
with wavelengths longer than the box size are missing, so the application of
our modelling method should be limited to
scales much smaller than the simulation box size. This is particularly true
for redshift-space distortion modelling, since the velocity field is more
sensitive to large-scale modes than the density field.

In presenting the method, the halo variable is adopted to be halo mass (or
characteristic velocity for the subhalo case) to build the tables. The
corresponding HOD/CLF model assumes that the statistical properties of
galaxies inside haloes only depend on halo mass, not on halo environment or
growth history. Clustering of haloes at fixed mass is found to depend on the
assembly history (a.k.a. assembly bias; e.g.
\citealt{Gao05,Wechsler06,Zhu06,Jing07}). There is room for the galaxy
content in haloes of fixed mass to depend on halo formation history, which
would affect galaxy clustering and HOD constraints
\citep[e.g.][]{Zentner14}, although no clear evidence is found in
hydrodynamic galaxy formation simulations \citep[e.g.][]{Berlind03} or galaxy
clustering measurements \citep[e.g.][]{Lin15}. As mentioned in
Section~\ref{sec:method}, the halo variable in our method is not necessarily
the halo mass. It can certainly be a set of variables, like halo mass plus a
variable characterizing halo formation history (e.g. halo concentration or
formation redshift). With tables built in terms of the set of variables, 
along with an HOD/CLF model depending on these variables, the simulation-based 
method works in the same way as presented in this paper. However, the
efficiency of the method drops sharply when including more halo variables.
The limitation is mainly set by the computation of the two-halo terms, 
where both the table size and computational time scale as $\mathcal{O}(N^2)$,
with $N$ the total number of bins in halo properties (e.g. with $N_1$ halo 
mass bins and $N_2$ halo formation time bins, $N=N_1N_2$). In practice, we
may be barely able to accomodate the case of two halo variables, by choosing
bin sizes to minmize the table size and computational cost without sacrificing 
the accuracy of the method. Before resorting to directly populating the 
simulations, a possible way of circumventing the limitation is to use some 
combination of halo variables, reducing the problem to one effective halo 
variable. Certainly further investigations are needed to find the appropriate 
combination(s).

A different approach to model galaxy clustering is through an emulator 
\citep[e.g.][]{Kwan15}. With this approach, galaxy correlation functions are 
first obtained with mock catalogs from $N$-body simulations, spanning a range 
of HOD parameters. Then the emulator works by interpolation to predict the 
galaxy correlation function for any given set of HOD parameters. Compared to
the method we propose in this paper, the emulator can be extremely fast, 
since it only performs interpolations and avoids any calculation at the level 
of dark matter haloes. In principle, the emulator can be generalized to 
interpolate among the one-halo and two-halo component contributions to the 
2PCFs. However, by construction, the emulator only operates with a certain HOD 
form and within a certain range of HOD parameters for the interpolation to
work and for the accuracy to be under control. The method we propose performs
direct calculations with clear physical meanings based on halo properties, 
and therefore it does not suffer from the above restrictions of an emulator.

With increasingly more precise measurements of galaxy clustering from
forthcoming large galaxy surveys, such as DESI \citep{DESI} and Euclid
\citep{Euclid}, we expect that the
accurate and efficient modelling method introduced in this work and its
generalizations will have great potentials and wide applications.

\section*{Acknowledgments}
ZZ is partially supported by NSF grant AST-1208891. HG acknowledges the
support of NSFC-11543003 and the 100 Talents Program of the Chinese Academy of Sciences.

\appendix
\def\Vm{V_{\rm m}}
\def\Vo{V_{\rm o}}
\def\FVoi{F_{{\rm o},i}^{V_{\rm o}}}
\def\FVoj{F_{{\rm o},j}^{V_{\rm o}}}
\def\Foi{F_{{\rm o},i}}
\def\Foj{F_{{\rm o},j}}
\def\FVmi{F_{{\rm m},i}^{V_{\rm m}}}
\def\FVmj{F_{{\rm m},j}^{V_{\rm m}}}
\def\Fmi{F_{{\rm m},i}}
\def\Fmj{F_{{\rm m},j}}

\section{Covariance Matrix with Model Uncertainty}
\label{sec:appendix} Let us consider the case that we use a model built on
one simulation in a volume $\Vm$ (`m' for model) to interpret the
observation obtained from a survey volume $\Vo$ (`o' for observation).
What covariance matrix should we use to model the data? The covariance matrix
estimated for the observation tells us the covariance in the observational
data. However, the model is based on a simulation with a finite volume, and
therefore it is not the global model or the model from ensemble average. The
model itself has uncertainty, and the modelling needs to account for this.
To derive the effective covariance matrix $\mathbf{C}^{\rm eff}$ to be used
in the modelling, let us define the $i$-th data point measured in the
observational volume $\Vo$ as $\FVoi$, the $i$-th data point from the model
with simulation volume $\Vm$ as $\FVmi$, and the global averages (or the
ensemble averages) of the observational and model data points as $\Foi$
and $\Fmi$, respectively. Note that for an accurate model that reflects the
reality, we have $\Fmi = \Foi$. That is, the global model reproduces the
global average observation.

The effective covariance matrix with model uncertainty included is then
\begin{eqnarray}
C_{ij}^{\rm eff}
& = &
\left\langle
\left( \FVoi-\FVmi \right) \left( \FVoj-\FVmj \right)
\right\rangle \label{eqn:1}\\
& = &
\left\langle
\left[ \left(\FVoi-\Foi\right) - \left(\FVmi-\Fmi\right) \right]\right.
\nonumber \\
& &
\,\,\left.\left[ \left(\FVoj-\Foj\right) - \left(\FVmj-\Fmj\right) \right]
\right\rangle \label{eqn:2}\\
& = &
\left\langle
\left(\FVoi-\Foi\right) \left(\FVoj-\Foj\right)
\right\rangle \nonumber\\
& & +
\left\langle
\left(\FVmi-\Fmi\right) \left(\FVmj-\Fmj\right)
\right\rangle
\nonumber\label{eqn:3}\\
& &
+
\left\langle
\left(\FVoi-\Foi\right) \left(\FVmj-\Fmj\right)
\right\rangle \nonumber\\
& &
+
\left\langle
\left(\FVmi-\Fmi\right) \left(\FVoj-\Foj\right)
\right\rangle .
\end{eqnarray}
The symbol $\langle \rangle$ denotes global/ensemble average over
observations in volumes of $\Vo$ and over models in volumes of $\Vm$. From
(\ref{eqn:1}) to (\ref{eqn:2}), we make use of the above $\Fmi = \Foi$
relation. In (\ref{eqn:3}), the first term
is the element $C_{ij}^{\Vo}$ of the covariance matrix for the measurements
in volume $\Vo$, the second term is the element $C_{ij}^{\Vm}$ of the
covariance matrix for the measurements in volume $\Vm$ (since the model
values can be regarded as mock measurements), and both the third and fourth
terms are zero (since there is no correlation between observation
measurements and mock measurements). We then have
\begin{equation}
{\mathbf C}^{\rm eff} = {\mathbf C}^{\Vo} + {\mathbf C}^{\Vm},
\label{eqn:C_eff1}
\end{equation}
and the result is expected and intuitive.

For power spectrum or 2PCF measurements, the
covariance matrix element is inversely proportional to the volume
\citep{Feldman94,Tegmark97}. We can express the effective covariance matrix
in equation~(\ref{eqn:C_eff1}) in terms of the one estimated for the
observation and the relative volume of the simulation and observation,
\begin{equation}
{\mathbf C}^{\rm eff} = \left(1+\frac{\Vo}{\Vm}\right){\mathbf C}^{\Vo}.
\label{eqn:C_eff2}
\end{equation}

\end{document}